\documentclass[preprint,showpacs,preprintnumbers,amsmath,amssymb]{revtex4}

\usepackage{graphicx}
\usepackage{dcolumn}
\usepackage{bm}

\begin{document}

\preprint{}

\title{Measurement time in double quantum dots}

\author{H. Cruz}
 \email{hcruz@ull.es}
\affiliation{%
{\it Departamento de F\'\i sica B\'asica and IUdEA, }\\{\it Universidad de La Laguna,
38204 La Laguna, Tenerife, Spain.}}%
\date{\today}

\begin{abstract}
In this letter, we have considered an electron in a double quantum dot
system interacting with a detector represented by a point contact. We
present a dynamical model for the gradual decoherence of the density
matrix due to the interaction with the detector. The interaction of the 
qubit (quantum system) on the quantum point contact (environment) leads to a 
discrete set of pointer states of the double quantum dot (apparatus). The necessary time for
the qubit decoherence process (measurement time) has been calculated through 
this model. The 
existence of a minimum time for the quantum measurement has been obtained.
\end{abstract}

\maketitle

The problem of understanding whether a measurement process can be
analyzed within the quantum mechanical formalism has long been a difficult
unresolved issue in the foundations of quantum mechanics \cite{wz}. On one hand, the
quantum theory states that the vector corresponding to a physical system
undergoes a continuous evolution governed by Schr\"{o}dinger equation; on the
other hand, the theory prescribes a sudden jump motion to the state of a
physical system undergoing a measurement by an external device. Von
Neumann's projection rules \cite{von} are indeed to be added to the quantum
formalism in order to account for the transition from a pure to a mixed
state (the so-called wave function collapse), and this makes quantum
mechanics a non-self-contained theory. 

The renewed interest in the
measurement problem is justified by the development of mesoscopic systems
sensitive to the phase of the electronic wave function. Recent proposals
suggested using mesoscopic devices, such a Josephson junctions or coupled
quantum dots, as quantum bits (qubits), which are the basic elements of
quantum computers \cite{joseph}. Among various modern approaches to the
measurement problem in mesoscopic structures let us mention the idea of
replacing the collapse postulate by the gradual decoherence of the density
matrix due to the interaction with the detector \cite{zrmp}.
Decoherence is the emergence of classical features of a quantum system resulting from its interaction with
the environment. Zurek has proposed that the interaction of the quantum system on the environment
leads to a preferred and discrete set of quantum states (pointer states), which remain robust. This
environment selection of the preferred pointer states was termed einselection \cite{zprd}.
In the simplest models that have been used up to now to study decoherence, such pointer states are
eigenstates of the pointer observable which commutes with the system-environment interaction
Hamiltonian. These interaction Hamiltonians are generally based on spin-atom interaction and such a
concept can be generalized using the predictability sieve \cite{schloss}.

Solid-state quantum devices have been demonstrated as promising systems for quantum computation \cite{t1}.
Recently, Gurvitz \textit{et. al.} \cite{gurvitz} have considered a qubit
interacting with its environment and continuously monitored by a detector
represented by a point contact. In such a case, the decoherence rate $\Gamma
_{d}$ due to interaction with the detector is inversely proportional to the
measurement time $\Delta t$. For strong coupling to the detector, i.e., $%
\Gamma _{d}\rightarrow \infty $ and $\Delta t\rightarrow 0$, the measurement
is idealized to be instantaneous. Accordingly, the electron in the coupled
quantum dot system instantaneously makes the transition $|\psi
_{i}>\rightarrow |\psi _{f}>$ by measurement (the so-called quantum-jump 
\cite{milburn}). 
Controlled decoherence of electrons 
has been studied experimentally in recent years \cite{s1,f1}. 
Ferry {\it et. al.} \cite{f1} have shown that the conductance oscillations exhibited by open quantum dots
are governed by a discrete set of stable quantum states which have the properties of the pointer states.

An account of decoherence of a double quantum dot (a qubit), interacting with a measurement device, has
become a problem of crucial importance in quantum computing.
To build up a quantum computer using quantum dots, it is necessary to know the decoherence
time for a qubit when a measurement takes place. \cite{g2}
This is the aim of the present work. In this letter,
we present a dynamical model for the gradual decoherence of the density matrix due to
the interaction with the measurement device. 
In the model, the interaction of the electron (qubit) on the environment (quantum point contact) leads
to a discrete set of pointer states, which commute with the system-environment $H_{int}$ Hamiltonian.
The measurement times will be calculated through this model.

Let us now consider electrostatic quantum bit measurements (Fig. 1) in double quantum
dots \cite{meas}. The qubit is a single electron in a double quantum dot and the detector is a point
contact placed near one of the dots \cite{butt}. 
We have also considered a one-dimensional model (quantum wire) for the quantum point contact.
If the electron occupies
the first dot, the transmission coefficient of the point contact decreases
due to electrostatic repulsion generated by the electron (Fig. 2). Thus, the electron
position is monitored by the tunneling current. The qubit can be described
by the Hamiltonian \cite{gurvitz} $H=H_{DQD}+H_{QPC}+H_{int}$, where
\begin{equation}
H_{DQD}=\varepsilon_1 |A_1><A_1|+ \varepsilon_2 |A_2><A_2| \text{,}  \label{1}
\end{equation}%
is the double quantum dot Hamiltonian and $|A_{1,2}>$ are the double dot eigenstates (Fig. 2).
The term $H_{QPC}$ describe 
the quantum point contact detector
\begin{equation}
H_{QPC}= \sum_l \varepsilon_l |\psi_l><\psi_l|+ \sum_r \varepsilon_r |\psi_r><\psi_r|
+ \sum_{l,r} t_{lr} ( |\psi_l><\psi_r|+ |\psi_r><\psi_l|)
 \text{,}  \label{1}
\end{equation}%
where $| \psi_l>$ and $| \psi_r>$ are electron states in the left and right reservoirs, respectively, and 
$t_{lr}$ is the hopping amplitude between the states $| \psi_l>$ and $| \psi_r>$. The $H_{int}$ term is the interaction with the qubit, i.e., 
\begin{equation}
H_{int}= |A_1><A_1| \otimes  \sum_{l,r} \delta t_{lr} (|\psi_l><\psi_r|+ |\psi_r><\psi_l|)
 \text{.}  \label{1}
\end{equation}%

The interaction term
generates a change in the hopping amplitude $\delta t_{lr}=t'_{lr}-t_{lr}$ (Fig. 2).
We should point out that the interaction term is between each 
eigenstate in the quantum point contact and the electron in the double quantum dot.
Thus the detector current is I' (due to the hopping term $t'_{lr}$) when the electron occupies the first dot state $| A_1 >$, and I (due to $t_{lr}$) when the electron
occupies the second dot state $| A_2 >$ (Fig. 2). For simplicity we have considered electrons as spinless fermions and a constant
electric field in the double quantum dot system that is generated by the electrons in the quantum point 
contact. In addition to this, we have neglected
the effects of the applied bias voltage and the electron-electron interaction in the quantum point contact. 

In order to solve $H_{QPC}$, we have chosen a confining (infinite) potential in the left and right extremes
of the point contact (Fig. 2). In such a case, 
the eigenstates for a particle which is moving in a symmetric double quantum well
can be written  \cite{qtt} as
\begin{equation}
\varepsilon_{n}= \varepsilon_{n}^{qw} \pm \delta \varepsilon_{n} 
 \text{,}  \label{1}
\end{equation}%
where $ \varepsilon_{n}^{qw} $ is the quantum well eigenvalue and $ \delta \varepsilon_{n} $ is
given by \cite{qtt}
\begin{equation}
\delta \varepsilon_{n} \cong 2 \varepsilon_{n}^{qw} \frac{\sqrt{2 m^* (V- \varepsilon_{n}^{qw} )}}{V m^* (l_{a} + l_{b}/2 ) } exp 
\left[ - \frac{l_{b}}{ \hbar} \sqrt { 2 m^* (V- \varepsilon_{n}^{qw} )  } \right]
 \text{,}  \label{1}
\end{equation}%
being $m^*$ the electron effective mass, $V$ the barrier potential and $l_a$ and $l_b$ the quantum well width and barrier
width in the quantum point contact, respectively (Fig. 2). The $ \delta \varepsilon_{n} $ term splits the energy
levels of the quantum well. In obtaining the approximate solution of Eq. (5) we have assumed 
that the width of the well $l_a$ is large \cite{qtt}, i.e., 
$ l_a >> l_b$ in our case.  

To leave pointer states in the detector untouched, the interaction Hamiltonian should be a function of the pointer 
observable $ \widetilde{A}$ of the apparatus (double quantum dot). 
We can notice that $[H_{int}, \widetilde{A}]=0$ in our model. As a result, pointer states are easy to characterized and are
the $ |A_{1,2}> $ eigenstates of the double quantum dot.
Now we introduce the environment, represented by a boson
bath at zero temperature and interacting with the electron. 
First, consider the case in which the
electron wave function $| \phi >$ is not coupled to the detector, i.e., $| \phi > = | A_2 >$. Then, $H_{int}$=0 and the barrier potential takes the lowest value $V_2$ (Fig. 2). 
Now let us consider the interaction with the detector. In such a case, $| \phi > = | A_1 >$ and the barrier potential takes the highest value $V_1$.
The eigenstates $ | \psi_{n}^{\pm} >$ and eigenvalues $ \varepsilon^{\pm}_{n} $ in the electron reservoir can be obtained by solving $H_{QPC}$ at
different barrier heights $V_1$ and $V_2$ through Eq. (5). In our case, the energy level $\varepsilon^{+}_{n}$ ($\varepsilon^{-}_{n}$)
in the quantum point contact will correspond to $V_1$ ($V_2$).
We note that the interaction Hamiltonian can be easily diagonalized. In our model, both
{$| A_{1,2} >$} and {$ | \psi_{n}^{\pm} >$} are stationary states in the double
quantum dot and the quantum point contact, respectively. 
When the electron in the double dot is in the stationary state $| \phi > = | A_1 >$ ($| \phi > = | A_2 >$),
each electron in the quantum point contact occupies the stationary state 
$| \psi^{+}_{n} > $ ($| \psi^{-}_{n} > $). We notice that our Hamiltonian doesn't depend explicitly on time. 
In such a case, the initial state of the total hamiltonian \cite{zrmp}
\begin{equation}
| \Phi (0) >= a | A_1 > \left( \prod_{n} |\psi_{n}^{+}> \right) +  b | A_2 > \left( \prod_{n} |\psi_{n}^{-}>\right) \text{,}  \label{1}
\end{equation}%
where $a$ and $b$ are two normalization constants, evolves into
\begin{equation}
| \Phi (t) >= a e^{-\frac{i}{\hbar}\varepsilon_{1} t}|A_1>\left( \prod_{n} e^{-\frac{i}{\hbar}\varepsilon_{n}^{+} t} |\psi_{n}^{+}> \right)  +  b e^{-\frac{i}{\hbar}\varepsilon_{2}t}|A_2 >\left( \prod_{n} e^{-\frac{i}{\hbar}\varepsilon_{n}^{-}t} |\psi_{n}^{-}>\right) \text{,}  \label{1}
\end{equation}%
and the reduced density matrix for the apparatus \cite{zrmp} is
\begin{equation}
\rho_A = |a|^2 |A_1 > <A_1|+ab^* r(t)|A_1 > <A_2|+a^* b r^{*}(t) |A_2 > <A_1| +|b|^2 |A_2 > <A_2|  \text{,}  \label{1}
\end{equation}
being
\begin{equation}
r(t)= e^{-\frac{i}{\hbar}(\varepsilon_{1}-\varepsilon_{2}) t} \prod_{n} e^{-\frac{i}{\hbar}(\varepsilon_{n}^{+}-\varepsilon_{n}^{-}) t} <\psi_{n}^{-} |\psi_{n}^{+}>
 \label{1}
\end{equation}%
a coefficient that determines the relative size of the off-diagonal terms. We note that 
if each electron in the quantum point contact occupies an specific $a|\psi_{n}^{+}>+b|\psi_{n}^{-}>$ state
in energy space, the off-diagonal terms $r(t)$ will not vanish as time progresses. 

We shall now consider that each electron in the
quantum point contact will spread out its wave function towards a higher state due to thermal effects. 
For this purpose, we consider a wave packet which is localized in energy space around a quantum number
$\bar{n}$, that is, a coefficient $\epsilon $ assumes only significant values in a small vicinity of
$n= \bar{n}$. The approximation is valid for a narrow energy distribution of the wave packet.
Only
nearest neighbor states will be considered to simplify.
In such a case, the initial wave function is
\begin{equation}
| \Phi (0) >= a | A_1 > \left( \prod_{n=1}^{N-1} \left[ (1- \epsilon)|\psi_{n}^{+}>+ \epsilon |\psi_{n+1}^{+}> \right] \right) +  b | A_2 > \left( \prod_{n=1}^{N-1} \left[ (1- \epsilon)|\psi_{n}^{-}> + \epsilon |\psi_{n+1}^{-}> \right] \right) \text{,}  \label{1}
\end{equation}%
where $\epsilon$ is a small quantity
$|\epsilon|<<1$, that is also normalized, $|1- \epsilon|^2+|\epsilon|^2=1$, $N$ is the number of electrons in the double point contact and $r(t)$ is
\begin{equation}
r(t)= e^{-\frac{i}{\hbar}(\varepsilon_{1}-\varepsilon_{2}) t} \prod_{n=1}^{N-1} \left[ |1- \epsilon|^2 e^{-\frac{i}{\hbar}(\varepsilon_{n}^{+}-\varepsilon_{n}^{-}) t} + |\epsilon |^{2} e^{-\frac{i}{\hbar}(\varepsilon_{n+1}^{+}-\varepsilon_{n+1}^{-}) t} \right].
 \label{1}
\end{equation}%
To simplify our calculations, we have considered that the wave function forms in the double point contact are
slightly modified due to the barrier potential variation. Then, we have approached $<\psi_{n}^{-} |\psi_{n}^{+}> \sim 1$, $<\psi_{n+1}^{-} |\psi_{n+1}^{+}> \sim 1$, $<\psi_{n}^{-} |\psi_{n+1}^{+}> \sim 0$ and $<\psi_{n+1}^{-} |\psi_{n}^{+}> \sim 0$.
For large environments
consisting of many $N$ electrons at large times the off-diagonal terms will now vanish.
The results have been checked for more than one nearest neighbor state (and for
different $\epsilon$ values in the $|\psi_{n}^{+}>$ and $|\psi_{n}^{-}>$ states). In such cases, $r(t)$
vanished faster.

In the quantum point contact, we have
considered a GaAs/Ga$_{1-x}$Al$_{x}$As double quantum well system which
consists of two $l_a =5000$\AA   wide GaAs quantum wells separated by a barrier of
thickness $l_b =$100\AA\ (Fig. 2) . The barrier height and electron
effective mass are taken to be $V_2 =220$meV and 0.067$m_{0}$, respectively \cite{hc}. 
The $V_1$ value has been obtained considering the Coulomb potential generated by a single
electron in the $A_1$ quantum well. The distance between the $A_1$ quantum well and the barrier is taken to be
$100$\AA. In such a case, $V_2 - V_1 \sim 12$meV. 
Fig. 3 and 4 show  $ z(t) = |r(t) |^2 $ versus time at different $ N $ and $ \epsilon $ values,
respectively. 
It is found that the magnitude of the off-diagonal terms decreases exponentially fast, with
the physical size $ N $ of the environment effectively coupled to the state of the double quantum dot. 
As a result, the qubit interaction on the environment leads to two different pointer states $\left\{ |A_1> , |A_2>  \right\}$
of the double quantum dot, which remain robust in the einselection approach.
We note that the effectiveness of einselection depends on the initial state of the environment, $\epsilon$. When
the environment is in an eigenstate of $H_{int}$, the coherence in the system will be retained ($ \alpha= | \epsilon |^2 =0.001 \sim 0$ in Fig. 4). This
special environment state is, however, unlikely in realistic circumstances.

Finally, we think that it could be possible to monitoring the decoherence process. If 
the number of electrons in the point contact electrodes is not large enough, the electron
wave function in the double quantum dot will not collapse. The $ |\epsilon |^2$ parameter
could be controlled by decreasing the temperature. In such a case, the quantum measurement
will not be obtained. The number of injected electrons in the quantum point contact could be
controlled by external contacts.
From Fig. 3, we note that it is easy to define a measurement time. The $ |r(t) |^2 $ term
decreases exponentially fast with the number of electrons $N$. If $ |r(t) |^2 $ takes a value
such $ | r(t) |^2 < 0.1$ when $t> \tau$, we can define a measurement time $ \tau $. In such a case, we have
a 0.90 probability of wave function collapse. Fig. 5 shows measurement time versus $N$. The 
$ \tau $ value decreases exponentially with the number of electrons and finally reaches a near
constant value, $\tau \sim 175 ps$. It is found the existence of a minimum time for the
measurement process at a fixed $ \epsilon $ value. If an oscillating current is applied to the electrodes,
the measurement process will happen only if the current oscillation period is higher than the
measurement time. In this way, it could be also possible to monitoring the wave function decoherence
using an external oscillating current.

In summary, in this letter we have considered a qubit (electron) in a double quantum dot
interacting with a measurement device. We present a dynamical model for the gradual
decoherence of the density matrix due to the interaction with the quantum point contact.
The interaction of the qubit on the quantum point contact environment
leads to a discrete set of pointer states of the double quantum dot.
The neccesary time for the measurement has been calculated through this model. It
is found the existence of a minimum time for the qubit decoherence process due to
the interaction with the detector.

\newpage

\section{Figures and Tables}

\begin{itemize}
\item \textbf{FIG. 1} 
A schematic illustration of the measurement device. The
point contact detector with two separate reservoirs near the
double quantum dot. The left and right electrodes are filled up
with electrons in the $| \psi_l >$ and $| \psi_r >$ states,
respectively. the qubit is an electron in the double quantum dot
system, $| \phi >= a | A_1 > + b | A_2 > $.

\item \textbf{FIG. 2} A schematic illustration of the measurement device. The $x$ axis 
is set longitudinally. A point
contact detector monitoring the electron position in the double
quantum dot. If the electron wave function is not coupled to the
detector, $| \phi > = | A_2 > $, the barrier potential takes the
$V_2 $ value and $ \varepsilon_{n}^{-}$ are the electron eigenvalues
in the reservoir. When $| \phi > = | A_1 > $, the barrier potential 
takes the highest value $V_1$ and $ \varepsilon_{n}^{+}$ the electrons
in the quantum point contact.

\item \textbf{FIG. 3} 
$z(t)$ versus time at different number of electrons $N$ values. The $z$ term  is $z(t)=|r(t)|^2$
and  $ \alpha = | \epsilon |^2 = 0.1$.

\item \textbf{FIG. 4 } 
$z(t)$ versus time at different $ \alpha = | \epsilon |^2 $ values. The $z$ term  is $z(t)=|r(t)|^2$ 
and $N$=25. 

\item \textbf{FIG. 5 } Measurement time versus number of electrons $N$ at $ \alpha = | \epsilon |^2 = 0.1$.

\end{itemize}

\end{document}